\documentclass[12pt]{article}
\usepackage{epsf}
\usepackage{latexsym}
\newcommand{\be}{\begin{equation}}
\newcommand{\ee}{\end{equation}}
\newcommand{\ba}{\begin{array}}
\newcommand{\ea}{\end{array}}
\newcommand{\bqa}{\begin{eqnarray}}
\newcommand{\eqa}{\end{eqnarray}}
\begin{document}
\begin{center}

{\Large\sf The Use of Analyticity in the $\pi\pi$ and $K\bar K$
 Coupled Channel System\footnote{Summary of talks given by Z.X.
at $International$ $Conference$ $on$ $Flavor$ $Physics$ $2001$,
Zhangjiajie, Hunan, May 30th, 2001; given by H.Z. at BES annual
meeting,
 Jixian, Tianjin, June
 11th, 2001
 and at
Eurodaphne workshop on $Nonperturbative$ $Methods$ $in$ $Chiral$
$Theories$, Valencia, Spain, June 28th, 2001.}}
\\[10mm]
{\sc Zhiguang Xiao and Hanqing Zheng\footnote{e-mail:
zheng@th.phy.pku.edu.cn}}
\\[5mm]
{\it Department of Physics, Peking University, Beijing 100871, P.~R.~China
}
\\[5mm]
\begin{abstract}
Studies on the IJ=00 $\pi\pi$ and $K\bar K$ coupled--channel
system are made using newly derived dispersion relations between
the phase shifts and poles and  cuts. It is found that the
$\sigma$ resonance must be introduced to explain the experimental
phase shifts, after evaluating the cut contribution. The effects
of nearby branch point singularities to the determination of the
$f_0(980)$ resonance are also carefully clarified.
\end{abstract}
\end{center}

The revived interests in the $\sigma$ particle has stimulated many
investigations in recent years~\cite{recentsigma}. The difficulty
in answering the question whether  there exists the $\sigma $
particle comes from the fact that the $\sigma $ particle, if
exist, must be a very broad resonance as indicated by the broad
enhancement of the experimental phase shifts below 1GeV in the
IJ=00 channel. However it is difficult to distinguish the
contribution of a broad resonance from the contribution of the
left hand cut both theoretically and experimentally. In our  point
of view it is difficult to recognize the existence of the $\sigma$
meson before one can seriously estimate the left hand cut effects,
though the $\sigma $ resonance can be naturally generated from
many dynamical  models producing unitarized partial wave
amplitudes fit appropriately to the experimental data.

In two recent papers~\cite{xz1,xz2} the present authors devoted to the study on the
influence of  left hand cuts to the determination of the $\sigma $ meson.
For the purpose of separating the cut contribution from the pole contribution to the
$S$ matrix we developed
a  dispersion relation\footnote{Assuming however no bound state exists.} which reads,
\be\label{single}\label{main1}
\sin(2\delta_\pi)=\rho\left(\sum_i{{\rm Res}[{\cal
F}(z_i^{II})]\over s-z_i^{II}} +{1\over
\pi}\int_{-\infty}^0\frac{{\rm Im}_L{\cal F}}{s'-s}ds'
 + {1\over \pi}\int_{4m_K^2}^\infty\frac{{\rm Im}_R{\cal F}}{s'-s}ds'\right)\ ,
 \ee
where $\delta_\pi$ is the $\pi\pi$ phase shift in the single
channel unitarity region, and $z_i^{II}$ denotes the  resonance
pole position on the complex $s$ plane. The function ${\cal F}$ is
the analytic continuation of twice of the real part of the
$\pi\pi$ scattering T matrix (defined in the physical region) on
the complex $s$ plane, or in short-hand notation, ${\cal F} =2{\rm
Re}_R {\rm T}_{\pi\pi}$ in the physical region, and $\rho$ is the
kinematic factor, $\rho(s)=\sqrt{1-4m_\pi^2/s}$. Other relations
which are helpful in understanding Eq.~(\ref{main1}) are, \be
\sin(2\delta_\pi)=\rho{\cal F}\ ,\,\,\, {\rm Res}\left[{\cal
F}(z_i) \right]=\frac{i}{2\rho(z_i) S'(z_i)}\ . \ee The right hand
integral appeared on the $r.h.s.$ of Eq.~(\ref{main1}) is absent
in the single channel approximation. In the $\pi\pi, \bar K K$
coupled channel case one has \be\label{ImRF} {\rm Im}_R{\cal
F}=(1/\eta -\eta )\cos(2\delta_\pi)\ , \ee
 and hence the right hand
integral in Eq.~(\ref{main1}) can be evaluated using  available
experimental data at higher energies. However its contribution is
found to be very small even though the right hand integral
develops a cusp structure below the $\bar KK$ threshold. In
Eq.~(\ref{ImRF}) the definition of $\delta_\pi$ and $\eta$ comes
from the standard parametrization of the $S$ matrix in the coupled
channel unitarity region, \be\label{S}
 S=\left( \matrix{ {\eta {e^{ 2i\delta_{\pi}}}}, &
{i\sqrt{ 1-\eta^2}{e^{i\left(\delta_{\pi}+\delta_K\right)}}}\cr
{i\sqrt{ 1-\eta^2}{e^{i\left(\delta_{\pi}+\delta_K\right)}}}, &
{\eta {e^{ 2i\delta_K}}}\cr}\right)\ .
 \ee
It is important to realize that the $\pi\pi$ phase shift appeared
in Eq.~(\ref{S}) and Eq.~(\ref{ImRF}) is defined in the coupled
channel region and is different from the $\delta_\pi$ appeared in
Eq.~(\ref{main1}). Before going to the coupled channel case let us
focus on Eq.~(\ref{main1}) for the purpose of clarifying the
essential characters of the $\sigma$ resonance in a simpler way.
Our analysis revealed that neglecting the effects of the $\bar KK$
threshold and the $f_0(980)$ narrow resonance does no harm to the
qualitative understanding of the $\sigma $ meson~\cite{EO}.

The simplified version of Eq.~(\ref{main1}) , i.e., without the right
hand integral and the $f_0^{II}(980)$ resonance looks like the following,
\begin{equation}  \label{FL2}
\sin(2\delta_\pi(s))=\rho\left(\sum_{z_i=z_\sigma,z_\sigma^*}
{\frac{i/2\rho(z_i) }{S^{\prime}(z_i)(s-z_i)}} +
a+{\frac{s-m_\pi^2/2}{\pi }}\int_{-\infty}^0
{\frac{{\rm Im}_L F }{(s'-m_\pi^2/ 2)(s^{\prime}-s)}}ds^{\prime}\right)\ \ .
\end{equation}
In Eq.~(\ref{FL2}) we have on the $l.h.s.$ the experimental data
of $\sin(2\delta_\pi)$  truncated at certain scale below  the
$\bar KK$ threshold in order to exclude the threshold and the
narrow $f_0^{II}$ effects. On the $r.h.s.$ we have the $\sigma$
pole term put by hand hence one has to demonstrate that such a
term is really needed. However in order to make such a
demonstration possible one has to be able to estimate the left
hand integral in a reliable way, at least in the qualitative
sense. The integral is once subtracted at the point which happens
to be the Adler zero of the IJ=00 lowest order  chiral amplitude.
The subtraction  constant, $a$,  is not fixed by dispersion
theory. The way we estimate the left hand integral is to use the
1--loop chiral perturbation theory (CHPT) result on ${\rm
Im}_L{\cal F}$ and truncate the integral at certain scale
$\Lambda$ above which CHPT results become no longer trustworthy.
Implicit in our approximation  is the assumption that physics
above the scale $\Lambda$ does not influence the low energy
physics at qualitative level.  One may argue that the major
contribution from  high energies is already included in the
subtraction constant which has to be determined   by the fit,
rather than the theoretical calculation  here. We also make use of
 the [1,1] Pad\'e approximant
of CHPT to estimate the left hand integral, as inspired by a
series of recent efforts~\cite{oller}. Of course, the use of the
unitarized amplitude automatically regularize  the high energy
contribution to the left hand integral.  We bear in mind that none
of these estimates is perfect in the eye of a perfectionist.

Having  estimated  the left hand integral contribution to the
physical observable, $\sin(2\delta_\pi)$, it becomes possible now
to address  the problem whether the  $\sigma$ resonance is really
needed. The situation is nicely summarized in fig.~1 which is a
modified version of fig.~6 in Ref.~\cite{xz1}, that is in fig.~1
the effect of the kinematic factor $\rho$ is removed. It is
clearly shown in fig.~1 that the experimental data exhibit a very
broad peak or an enhancement   which can not be explained by the
dynamical cut effects: the latter is concave irrespective of the
different choice of the cutoff parameter. Actually, the derivative
of the cut contribution to $\cal F$ with respect to $s$ take the
following form, \be {d\over ds}{\cal F}|_{cut}={1\over
\pi}\int_{-\Lambda^2}^0 { {\rm Im}_L{\cal F}\over (s'-s)^2}ds'\
,\,\,\, {d^2\over ds^2}{\cal F}|_{cut}={2\over
\pi}\int_{-\Lambda^2}^0 { {\rm Im}_L{\cal F}\over (s'-s)^3}ds'\  .
\ee Since both the CHPT prediction and the [1,1]  Pad\'e
prediction on the sign of ${\rm Im}_L{\cal F}$ are always negative
within  a reasonable range of the $\Lambda $ parameter, it is
clear from the above expressions that the cut contribution to
${\cal F}$ must be concave.

\begin{figure}[hbtp]
\begin{center}
\vspace*{-0mm} \mbox{\epsfysize=90mm\epsffile{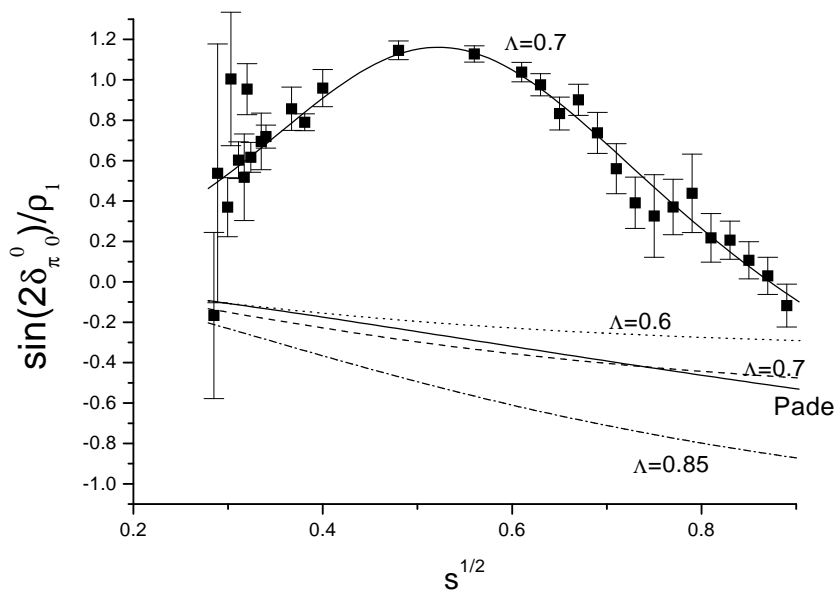}}
\vspace*{0mm} \caption{\label{fig00}A typical fit of 5 parameters
(4 resonance + 1 subtraction constant) in the I=J=0 channel, with
$\Lambda_{\chi PT}$=0.7GeV.
Different estimates on the left--hand integral are also plotted: The
dotted line corresponds to $\Lambda_{\chi PT}=0.6 GeV$, the dashed line corresponds to
$\Lambda_{\chi PT}=0.7 GeV$, the dot--dashed line corresponds to $\Lambda_{\chi PT}=0.85 GeV$
and the solid line corresponds to the Pad\'e solution.}
\end{center}
\end{figure}

Therefore it becomes unavoidable to call for the $\sigma$ resonance which turns over
the curve in fig.~1.
In the fit by using Eq.~(\ref{FL2}), the derivative of the $S$ matrix at the pole
position is taken as a free parameter.
The uncertainty of our fit result at quantitative level
mainly comes from the uncertainty in the estimation of
the left hand cut contribution.  However, one can still
manage to determine the location of the $\sigma $ pole within a reasonable range,
by varying the $\Lambda $ parameter.
It is worth noticing that the inclusion of the left hand cut
drives the $\sigma$ pole moving towards
left on the complex $s$ plane~\cite{AN}, though the effect is not very strong.
In the fit we find that the recent $K_{e4}$ data from the E865
collaboration~\cite{E865} is
crucial in reducing the magnitude
of the scattering length
parameter towards the chiral result~\cite{CGL}.  But we point out here that the
global fit still favors a somewhat larger value
of the scattering length parameter $a_0^0$.

In order to discuss the properties of
another interesting resonance named $f_0(980)$, it is appropriate
to go to the coupled channel region. In such case we can also write down a series of dispersion
relation which are similar to Eq.~(\ref{main1}),\footnote{The function ${\bf \Phi}$ defined
in the following differs by a sign to that in Ref.~\cite{xz2}.}
 \bqa
\label{MAIN}
 &&(\eta+{1\over \eta}){\sin
(2\delta_\pi)}= \rho_1\left(\Phi_{11}(s)+{1\over
2\pi i}\int_L {Disc\left({\bf TC}\right)_{11}\over z-s}dz\right)\ ,\nonumber\\
&&(\eta+{1\over \eta}){\sin (2\delta_K)}=  \rho_2\left(\Phi_{22}(s)+{1\over 2\pi i}\int_L
{Disc\left({\bf TC}\right)_{22}\over z-s} dz \right)\ ,\nonumber\\
&&\sqrt{1-\eta^2}(\cos(\delta_\pi+\delta_K)+{1\over
\eta}\cos(\delta_\pi-\delta_K))= {\sqrt{\rho_1\rho_2}}
\left(\Phi_{12}(s)+{1\over 2\pi i}\int_L
{Disc\left({\bf TC}\right)_{12}\over z-s} dz \right)\nonumber\\
\eqa in which the matrix function {\bf TC} is the sum of the T
matrices defined on different sheets:
 \be\label{C}
 {\bf TC}(z) \equiv {\rm T^{I}(z)+T^{II}(z)+T^{II}(z)+T^{IV}(z)} .
 \ee
In Eq.~(\ref{MAIN}) the matrix function ${\bf \Phi}$ represents
the sum over all possible pole contributions on different sheets.
The left hand integrals appeared in Eq.~(\ref{MAIN})  reflects the
effect of the left hand cut in ${\rm T}_{\bar KK}$ generated by
$t$ channel $2\pi$ exchanges: it starts from $4m_K^2-4m_\pi^2$ to
$-\infty$. We can immediately  draw some important conclusions by
comparing Eq.~(\ref{main1}) with Eq.~(\ref{MAIN}):

\begin{enumerate}

\item The experimental data below the
$\bar KK$ threshold only contribute to the determination of the
second sheet pole, only the data above the second threshold
contribute to the determination of the 3rd and/or 4th sheet pole.
This simple observation explains the reason why the third sheet
pole found from various fits differ so much in the literature
whereas the results on the second sheet $f_0^{II}(980)$ narrow
resonance agree with each other qualitatively: the second sheet
pole is much easier to  fix by experimental data.
 Especially for
$f_0^{II}(980)$, it is almost uniquely determined by the data
which are very close but below the
$\bar K K$ threshold.

\item  The fit above the $\bar KK$ threshold
will be  polluted by the uncertainty of the left hand cut at
$(-\infty, 4m_K^2-4m_\pi^2]$,  but the cut is expected to be
smooth function in the absence of nearby narrow 3rd or 4th sheet
poles. Especially the cut influence is not important  and may be
neglected when only discussing the second sheet poles. A typical
dynamical assumption made in dynamical models like in the case of
using Lippman--Schwinger
 equation
is that one only takes care of the $s$ channel force and neglects
the crossed channel forces by assuming them to behave mildly.
Similar assumptions occurred  in the more phenomenological $K$
matrix fits. In some sense  our above analysis justifies the
commonly used  assumption for neglecting the left hand cut
provided that there is no narrow 3rd or 4th sheet pole close to
the branch point $4m_K^2-4m_\pi^2$. The latter condition seems to
be indeed satisfied in the IJ=00 $\pi\pi$ and $\bar KK$ coupled
channel system.
\end{enumerate}

To end the discussion we quote the estimated value of the pole
positions of the $\sigma$ and $f_0^{II}(980)$ resonances, which
may be a little bit optimistic:\footnote{Firstly the $\sigma$ pole
position is rather sensitive to $a_0^0$, secondly in obtaining the
results we fix the left hand integral ($-\infty,0]$ by the Pad\'e
solution.} \bqa &M_\sigma&=478\sim 500{\rm MeV}\ ,\,\,\,
\Gamma_\sigma=480\sim 550{\rm MeV}\ ;\nonumber \\
&M_{f_0^{II}}&=982\sim 990MeV\ , \,\,\, \Gamma_{f_0^{II}}=35\sim
37{\rm MeV}\ . \eqa

\vspace{0.5cm} \noindent $Acknowledgement$:  In studying  the
topics related to this talk, we have benefited much from
conversations with Profs. Milan Locher, Chuanrong Wang, Bingsong
Zou and especially Valeri Markushin. We would also like to thank
the organizers of the BES 2001 annual meeting,  the flavor 2001
conference and the Eurodaphne workshop on ``Nonperturbative
Methods in Chiral Theories" for their warm hospitality and
support.


\begin{thebibliography}{99}
\bibitem{recentsigma} Review of Particle Physics,
Eur. Phys. J. C15 (2000) 1. See also the proceedings of
YITP Workshop on Possible Existence of the sigma Meson and
Its Implications to Hadron Physics, Kyoto 2000.
\bibitem{xz1} Z.~G. Xiao and H.~Q. Zheng, hep-ph/0011260, to appear in Nucl. Phys. A.
\bibitem{xz2} Z.~G. Xiao and H.~Q. Zheng, hep-ph/0103042.
\bibitem{EO}See also, A.~Dobado and J.~R.~Pelaez, Phys. Rev. {\bf D56} (1997) 3057;
J.~A.~Oller, E.~Oset and J.~R.~Pelaez, Phys. Rev. {\bf D59} (1999)074001;
Erratum: ibid, {\bf D60} (1999) 099906.
\bibitem{oller}J.~A.~Oller, E.~Oset  and A.~Ramos, Prog. Part. Nucl. Phys. {\bf 45}
(2000) 157, and references therein.
\bibitem{AN}V.~V.~Anisovich, V.~A. Nikonov,  Eur. Phys.
               J. {\bf A8} (2000) 401.
\bibitem{E865}P.~Truoel $et$ $al.$ (E865 Collaboration),
hep-ex/0012012. Other data we use in the fit come from:   L.~Rosselet {\it et al.},
Phys. Rev. {\bf D15} (1977) 574;  D.~Cohen, Phys. Rev. {\bf D22},
2595(1980);  A.~D.~Martin and E.~N.~Ozmuth, Nucl.
Phys. {\bf B158}, 520(1979); G.~Grayer et al.,
Proc. 3rd Philadelphia Conf. on Experimental Meson Spectroscopy,
Philadelphia, 1972 (American Institute of Physics, New York, 1972)
5;  B.~Hyams et al., Nucl. Phys. B64, 134(1973).
\bibitem{CGL}G.~Colangelo, J.~Gasser and H.~Leutwyler, hep-ph/0103088.

\end{thebibliography}
\end{document}